# SeeReader: An (Almost) Eyes-Free Mobile Rich Document Viewer

**Scott CARTER, Laurent DENOUE**

**FX Palo Alto Laboratory, Inc.**
**3400 Hillview Ave., Bldg. 4**
**Palo Alto, CA 94304**
**{carter,denoue}@fxpal.com**

### Abstract

Reading documents on mobile devices is challenging. Not only are screens small and difficult to read, but also navigating an environment using limited visual attention can be difficult and potentially dangerous. Reading content aloud using text-to-speech (TTS) processing can mitigate these problems, but only for content that does not include rich visual information. In this paper, we introduce a new technique, SeeReader, that combines TTS with automatic content recognition and document presentation control that allows users to listen to documents while also being notified of important visual content. Together, these services allow users to read rich documents on mobile devices while maintaining awareness of their visual environment.

## 1. Introduction

Reading documents on-the-go can be difficult. As previous studies have shown, mobile users have limited stretches of attention during which they can devote their full attention to their device [8]. Furthermore, studies have shown that listening to documents can improve users' ability to navigate real world obstacles [11]. However, while solutions exist for unstructured text, these approaches do not support the figures, pictures, tables, callouts, footnotes, etc., that might appear in rich documents.

SeeReader is the first mobile document reader to support rich documents by combining the affordances of visual document reading with auditory speech playback and eyes-free navigation. Traditional eReaders have been either purely visual or purely auditory, with the auditory readers reading back unstructured text. SeeReader supports eyes-free structured document browsing and reading as well as automatic panning to and notification of crucial visual components. For example, while reading the text "as shown in Figure 2" aloud to the user the visual display automatically frames Figure 2 in the document. While this is most useful for document figures, any textual reference can be used to change the visual display, including footnotes, references to other sections, etc.

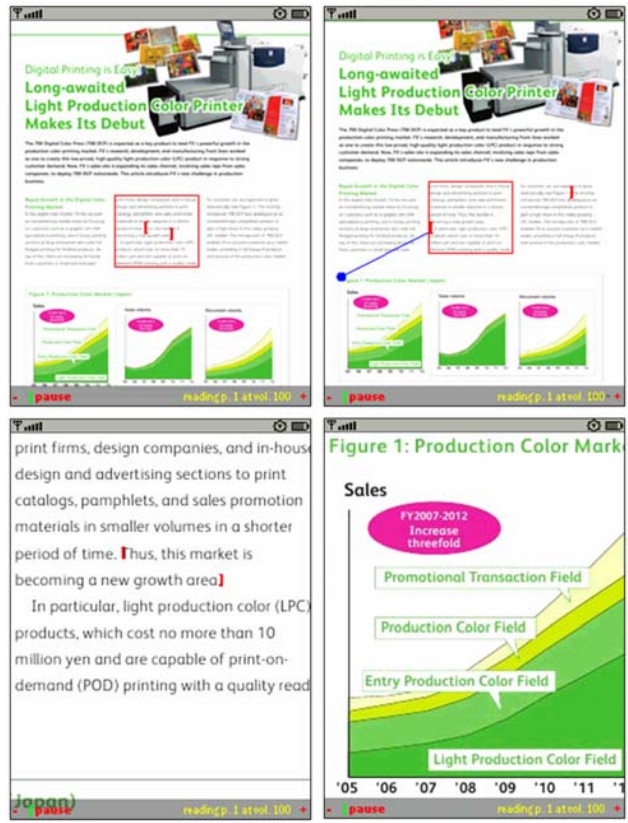

Figure 1: SeeReader automatically indicates areas of visual interest while reading document text aloud. The visual display shows the current reading position (left) before it encounters a link (right). When viewing the whole page, SeeReader indicates the link (top). When viewing the text, SeeReader automatically pans to the linked region (bottom). In both views, as the link is encountered, the application signals the user by vibrating the device.

Furthermore, SeeReader can be applied to an array of document types, including digital documents, scanned documents, and web pages. In addition to using references in the text to automatically pan and zoom to areas of a page, SeeReader can provide other services automatically, such as following links in web pages or initiating embedded macros. The technology can also interleave useful information into the audio stream. For example, for channel to free visual attention to the primary task. Along scanned documents the technology can indicate in the audio stream (with a short blip or explanation) when recognition errors would likely make text-to-speech (TTS) translation unusable. These indications can be presented in advance to allow users to avoid listening to garbled speech.

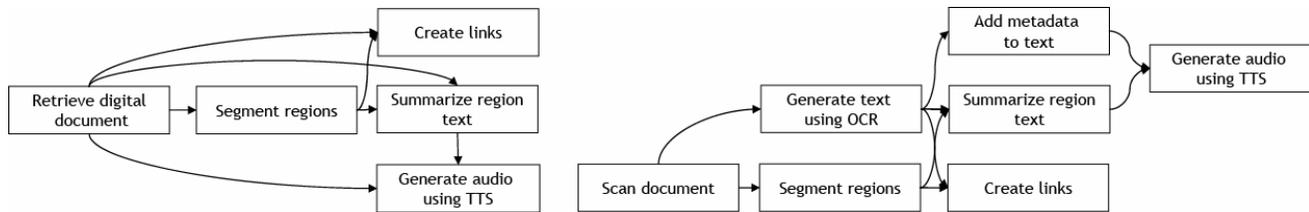

Figure 2: Data flow for digital documents (left) and scanned documents (right).

In the remainder of this paper, we briefly review other mobile document reading technologies, describe SeeReader including server processes and the mobile interface, and describe a study we ran to verify the usefulness of our approach.

## 2. Mobile Document Reading

The linear, continuous reading of single documents by people on their own is an unrealistic characterization of how people read in the course of their daily work. [1]

Work-related reading is a misnomer. Most "reading" involves an array of activities, often driven by some well-defined goal, and can include skimming, searching, cross-referencing, or annotating. For example, a lawyer might browse a collection of discovery documents in order to find where a defendant was on the night of October 3, 1999, annotate that document, cross-reference it with another document describing conflicting information from a witness, and begin a search for other related documents.

A growing number of mobile document reading platforms are being developed to support these activities, including specialized devices such as the Amazon Kindle[TM] (and others [10]) as well as applications for common mobile platforms such as the Adobe Reader[TM] for mobile devices. Past research has primarily focused on active reading tasks, in which the user is fully engaged with a document [9, 4]. In these cases, support for annotation, editing, and summarization is critical.

Our goal, on the other hand, is to support casual reading tasks for users who are engaged in another activity. A straightforward approach for this case is to use the audio these lines, the Amazon Kindle[TM] includes a TTS feature. However, the Kindle's provides no visual feedback while reading a document aloud. Similarly, the knfbReader[TM] converts words in a printed document to speech (http://www.knfbreader.com/). However, as this application was designed primarily for blind users, its use of the mobile device's visual display is limited to an indication only of the text currently being read. Other mobile screen readers, such as Mobile Speak (http://www.codefactory.es/), can be configured to announce when they have reached a figure, however as they do not link to textual references they are therefore likely to interrupt the reading flow of text. Similarly, with Click-Through Navigation [3] users can click on figure references in body text to open a window displaying the figure.

SeeReader improves on these techniques by making figures (and other document elements) visible on the screen automatically when references to them are read.

## 3. Analysis Pipeline

The SeeReader mobile interface depends upon third-party services to generate the necessary metadata. Overall, SeeReader requires the original document, information delineating regions in the document (figures, tables, and paragraph boundaries) as well as keyphrase summaries of those regions, the document text parsed to include links and notifications, and audio files. In this section, we describe this process (shown in Figure 2 in detail).

Initially, documents, whether they are scanned images or electronic (PDFs), are sent to a page layout service that produces a set of rectangular subregions based on the underlying content. Subregions might include figures and paragraph boundaries. In our approach, we use a version of the page segmentation and region classification described in [5]. Region metadata is stored in a database.

In the next step the body text is digitized. This is obviously automatic for electronic documents, while for scanned documents we use Microsoft Office[TM] OCR. Next, text is sent to a service that extracts keyphrases summarizing each region. Many text summary tools would suffice for this step. We use a version of [6] modified to work on document regions. Once processed, keyphrases are saved in a database.

Simultaneously, text is sent to an algorithm we developed to link phrases to other parts of the document. For example, our algorithm links text such as "Figure" to the figure proximate to a caption starting with text "Figure 2." Our algorithm currently only creates links with low ambiguity, including figures, references, and section headings using simple rules based on region proximity (similar to [7]). These links are also saved in a database.

Finally, the document text and region keyphrases are sent to a TTS service (AT&T's Natural Voices™). In the case of scanned documents, OCR scores are used to inject notifications to the user of potentially poorly analyzed blocks of text (e.g., this process may inject the phrase "Warning, upcoming TTS may be unintelligible"). The resulting files are processed into low footprint Adaptive Multi-Rate (AMR) files and saved in a database.

## 4. Mobile Application

The SeeReader mobile interface is a J2ME application capable of running on a variety of platforms. SeeReader also supports both touchscreen and standard input. Thus far, we have tested the device with Symbian devices (Nokia N95), Windows Mobile devices (HTC Touch), and others (e.g., LG Prada).

The application acquires documents and metadata and their associated metadata (as produced from the pipeline described above) via a remote server whose location is specified in a configuration file. The application can also read document data from local files. When configured to interact with a remote server, the application downloads data progressively, obtaining first for each document only XML metadata and small thumbnail representations. When the user selects a document (described below), the application retrieves compressed page images first, and AMR files as a background process, allowing users to view a document quickly.

The interface supports primarily three different views: document, page, and text. The document view presents thumbnail representations of all available documents. Users can select the document they wish to read via either the number pad or by directly pressing on the screen.

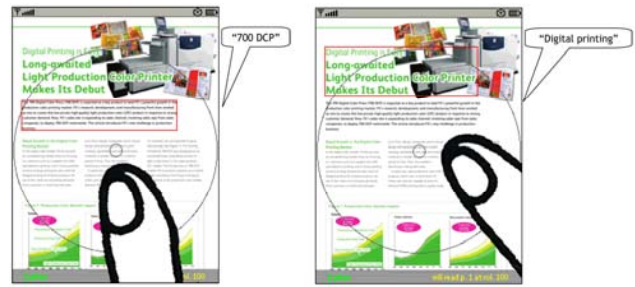

Figure 3: A user interacting with the touchwheel. As the user drags her finger around the center of the screen, the application vibrates the device to signal sentence boundaries and plays an audio clip of the keyphrase summarizing each region as it is entered.

After selecting a document, the interface is in page view mode (see Figure 1, top). We support both standard and touch-based navigation in page mode. For devices without a touchscreen, users can press command keys to navigate between regions and pages. For devices with a touchscreen, a user can set the cursor position for reading by pressing and holding on the area of interest. After a short delay, the application highlights the region the user selected and then begins audio playback beginning with the first sentence of the selected region.

To support *eyes-free* navigation, we implemented a modified version of the touchwheel described by Zhao et al. in [12] that provides haptic and auditory feedback to users as they navigate. This allows the user to maintain their visual attention on another task while still perusing the document. As the user gestures in a circular motion (see Figure 3), the application vibrates the device to signal sentence boundaries to the user. The application also reads aloud the keyphrase summary of each region as it is entered. In addition, we inject other notifications to help the user maintain an understanding of their position in the document as they navigate the touchwheel, such as page and document boundaries.

While in page view, users can flick their finger across the screen or use a command key to navigate between pages. Users can also click a command key or double-click on the screen to zoom in to the text view (see Figure 1, bottom). The text view shows the details of the current document page — users can navigate the page using either touch or arrow keys. Double-clicking again zooms the display back to the page view.

Multiple actions launch the read-aloud feature. In page view mode, when a user releases the touchwheel, selects a region by pressing and holding, or presses the selection key, the application automatically begins reading at the selected portion of text. The user can also press a command key at any time to start or pause reading. While reading, SeeReader indicates the boundaries of the sentence being read. When SeeReader encounters a link, it vibrates the device and either highlights the link or automatically pans to the location of the linked content, depending on whether the device is in page view or text view mode respectively (see Figure 1).

## 5. Evaluation

We ran a within subjects, dual-task study as a preliminary evaluation of the core features of the SeeReader interface. Participants completed two reading tasks while also doing a physical navigation task. Common dual-task approaches to evaluating mobile applications involve participants walking on treadmills or along taped paths while completing a task on the mobile device. These approaches are designed to simulate the bodily motion of walking [2]. However, they do not simulate the dynamism of a real-world environment. We developed a different approach for this study that focuses on collision avoidance rather than walking per se.

In this configuration, participants move laterally either to hit targets (such as doorways or stairs) or avoid obstacles (such as telephone poles or crowds). We simulated the targets and obstacles on a large display visible in the participant's periphery as they used the mobile device (see Figure 4). To sense lateral motion, a Wiimote™ mounted on the top of the display tracked an IR LED attached to a hat worn by each participant. We also included an override feature to allow a researcher to manually set the participant's location with a mouse in the event that the sensor failed. In addition to simulating a more dynamic situation, this approach has the advantage of being easily repeatable and producing straightforward measures of success for the peripheral task (task completion time and the number of barriers and targets hit).

In order to understand the benefits of eyes-free document reading, we compared the touchscreen SeeReader interface against a modified version of the touchscreen SeeReader interface with audio and vibration notifications removed (similar to a standard document reader). At the beginning of the experiment, we asked participants to open and freely navigate a test document using both interfaces. After 10 minutes of use, we then had participants begin the main tasks on a different document. We used a 2x2 design, having the participants complete two reading tasks, one on each interface, in randomized order. At the end of each task we asked participants two simple, multiple-choice comprehension and recall questions.

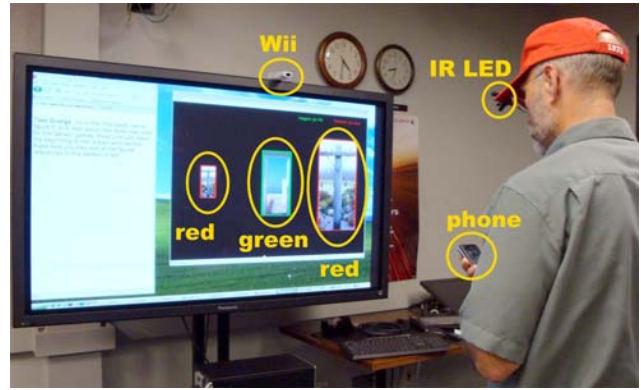

Figure 4: Study participants read documents while peripherally monitoring an interface, moving their body either to avoid barriers (red) or to acquire targets (green). A Wiimote™ tracked an IR LED attached to a hat to determine their location.

Finally, at the end of the study we asked participants the following questions (with responses recorded on a 7-point scale, where 1 mapped to agreement and 7 mapped to disagreement): "I found it easier to use the audio version than the non-audio version"; "I was comfortable using the touchwheel"; "It felt natural to listen to this document"; "The vibration notifications were a sufficient indication of links in the document". We ran a total of 8 participants (6 male and 2 female) with an average age of 38 (SD 6.02). Furthermore, only 2 of the 8 participants reported having extensive experience with gaming and mobile interfaces.

## 6. Results

Overall, we found that participants using SeeReader were able to avoid more barriers and hit more targets while sacrificing neither completion time nor comprehension of the reading material. Using SeeReader participants hit 76% (12% RSD) of the targets and avoided all but 10% (5% RSD) of the barriers, while using the non-audio reader participants hit 63% (11% RSD) of the targets and 17% (5% RSD) of the barriers. Meanwhile, average completion time across all tasks was virtually identical — 261 seconds (70 SD) for SeeReader and 260 seconds (70 SD) for the non-audio version. Also, comprehension scores were similar. Using SeeReader, participants answered on average 1.13 (.64 SD) questions correctly, versus 1 (.87 SD) using the non-audio version.

In the post-study questionnaire, participants reported that they found SeeReader easier to use (avg. 2.75, 1.58 SD), and felt it was natural to listen to the document (avg. 3.00, 1.31 SD). However, participants were mixed on the use of the touchwheel (avg. 4.38, 1.85 SD) and the vibration notifications (avg. 4.13, 1.89 SD).

While participants had some issues with SeeReader, overall they found it more satisfying than the non-audio version. More than one participant noted that they felt they did not yet have "enough training with [SeeReader's] touchwheel." Still, comments revealed that using the non-audio version while completing the peripheral task was "frustrating" and that some participants had no choice but to "stop to read the doc[ument]." On the other hand, SeeReader allowed participants to complete both tasks without feeling too overwhelmed. One participant noted that, "Normally I dislike being read to, but it seemed natural [with] SeeReader."

## 7. Discussion

The results, while preliminary, imply that even with only a few minutes familiarizing themselves with these new interaction techniques, participants may be able to read rich documents using SeeReader while also maintaining an awareness of their environment. Furthermore, the user population we chose was only moderately familiar with these types of interfaces. These results give us hope that after more experience with the application, users comfortable with smart devices would be able to use SeeReader not only to remain aware of their environment but also to read rich documents faster.

Methodologically, we were encouraged that participants were able to understand rapidly the peripheral task, and generally performed well. Still, it was clear that participants felt somewhat overwhelmed trying to complete two tasks at once, both with interfaces they had not yet encountered. We are considering how to iterate this method to make it more natural while still incorporating the dynamism of a realistic setting.

## 8. Conclusion

We presented a novel document reader that allows users to read rich documents while also maintaining an awareness of their physical environment. A dual-task study showed that users may be able to read documents with SeeReader as well as a standard mobile document reader while also being more aware of their environment.

In future work, we intend to experiment with this technique in automobile dashboard systems. We can take advantage of other sensors in automobiles to adjust the timing of display of visual content (e.g., visual content could be shown only when the car is idling). We anticipate that SeeReader may be even more useful in this scenario given the high cost of distraction for drivers.

## 9. Acknowledgements

We thank Dr. David Hilbert for his insights into improving the mobile interface. We also thank our study participants.

**Scott Carter** holds a PhD from The University of California, Berkeley. He is a research scientist at FX Palo Alto Laboratory.

**Laurent Denoue** holds a PhD from the University of Savoie, France. He is a senior research scientist at FX Palo Alto Laboratory.